\documentclass[prb,aps,twocolumn,preprintnumbers,superscriptaddress]{revtex4}
\usepackage{latexsym,epsfig,amssymb,amsfonts,amsmath,graphicx,color,bm,times,hyperref,amstext,epstopdf}

\bibpunct{[}{]}{;}{n}{}{}

 \usepackage{amsmath,amssymb,amsfonts,amstext}
\usepackage{setspace}                

\usepackage{graphicx,color,sidecap}          
\usepackage{epsfig}                  
\usepackage{wrapfig}
\usepackage{subfigure}
\usepackage{longtable}               
\usepackage{makeidx}                 
\usepackage{mathrsfs}                

\newcommand{\be}{\begin{equation}}
\newcommand{\ee}{\end{equation}}
\newcommand{\bea}{\begin{eqnarray}}
\newcommand{\eea}{\end{eqnarray}}

\usepackage{amsthm}
\theoremstyle{plain}

\usepackage{dsfont}                  
\usepackage{braket}
\usepackage{indentfirst,lipsum}

\begin{document}

\title{Spinon and bound state excitation ``light cones'' in Heisenberg XXZ Chains}
\author{A. L. de Paula Jr}
\affiliation{Departamento de F\'isica, Universidade Federal de Minas Gerais,
Belo Horizonte, MG, Brazil}
\author{H. Bragan\c{c}a}
\affiliation{Departamento de F\'isica, Universidade Federal de Minas Gerais,
Belo Horizonte, MG, Brazil}
\author{R. G. Pereira}
\affiliation{
International Institute of Physics, Universidade Federal do Rio Grande do Norte, 59078-970 Natal-RN,  Brazil, and \\
Departamento de F\'isica Te\'orica e Experimental, Universidade Federal do Rio Grande do Norte, 59072-970 Natal-RN, Brazil}
\author{R. C. Drumond}
\affiliation{Departamento de Matem\'atica, Universidade Federal de Minas Gerais,
Belo Horizonte, MG, Brazil}
\author{M. C. O. Aguiar}
\affiliation{Departamento de F\'isica, Universidade Federal de Minas Gerais,
Belo Horizonte, MG, Brazil}

\date{\today}

\begin{abstract}


We investigate the out-of-equilibrium dynamics  after a local quench that connects two spin-1/2 XXZ  chains prepared in the ground state of the Hamiltonian in different phases, one in the ferromagnetic phase and the other in the critical phase. We analyze the time evolution of the on-site magnetization and bipartite entanglement entropy via adaptive time-dependent density matrix renormalization group. In systems with short-range interactions, such as the one we consider, the velocity of information transfer is expected to be bounded, giving rise to a light-cone effect. 
Interestingly,  our results show that, when the anisotropy parameter of the critical chain is sufficiently close to that of the isotropic ferromagnet, the light cone is determined by the velocity of spin-wave bound states that  propagate faster than single-particle (``spinon'') excitations.  Furthermore, we investigate how the system approaches equilibrium in the inhomogeneous ground state of the connected system, in which the ferromagnetic chain induces a nonzero magnetization in the critical chain in the vicinity of  the interface.


\end{abstract}

\maketitle

\section{Introduction}

Non-equilibrium dynamics in strongly interacting systems has recently received considerable attention due to advances in numerical techniques \cite{dmrg1,dmrg2,dmrg3,dmrg4} and the possibility of simulating such systems in experiments with ultracold atoms in optical lattices \cite{optl2,optl3,optl4,optl5,optl6}. In this kind of experiments, the system  interacts rather weakly with the environment, thereby providing a unitary and coherent dynamics for long times. Due to high parameter control, such experiments enable one to drive the system to an out-of-equilibrium situation by means of quantum quenches \cite{optl1,sommer2011universal}, which can be either global or local. In both cases, a typical protocol is to initially prepare the system in an  eigenstate of a given Hamiltonian $ {H}_0$, then suddenly change some parameter, like magnetic field or interaction strength, and let the system evolve with the new Hamiltonian $ {H}$. In a global quench, a global parameter of the Hamiltonian, such as the magnetic
field acting on the whole system, is changed. This kind of quench is usually used to investigate questions about relaxation and thermalization  \cite{RigolPRA2006,CazalillaPRL2006,term1,term2,term3,term4,IlievskiPRL2015}. In the second case, the Hamiltonian is changed only locally - for example, a magnetic field can be switched on in part of the system. In this scenario, the non-equilibrium situation has been used to study the spread of energy, information, and correlations~\cite{local1,local2,local3,local4,SabettaPRB2013,BhaseenNP2015,BiellaPRB2016,de2014energy},
as well as transport properties~\cite{Bertiniarxiv2016}, the emergence of nonequilibrium steady states~\cite{local5}, and the thermal equilibration after the connection between two chains initially prepared at different temperatures~\cite{collura2014quantum, ponomarev2011thermal}. 

In this context, the seminal work of Lieb and Robinson \cite{lieb-robinson} is of relevance:
although non-relativistic Schr\"{o}dinger's equation imposes no limit on the speed, they showed that in many-body systems with short-range interactions the velocity of information propagation is bounded, leading to an effective light cone. This effect has  been confirmed in numerical studies \cite{local4,manmana2009time,bertini2016determination,lauchli2008spreading,carleo2014light,zamora2014splitting,gobert2005real,collura2015quantum,alba2014entanglement,vlijm2015quasi,jesenko2011finite} 
 and in experiments with ultracold atomic gases \cite{light-cone-exp1} and trapped ions \cite{light-cone-exp2}. The dependence of the light-cone effect on  system parameters, however, is still an open question. In this context, it has been shown that, in a global quench, the spreading velocities strongly depend on the temperature of the system through the initial density matrix~\cite{light-cone}. 
Within a semiclassical picture \cite{igloiprl,igloiprb}, supported by   conformal field theory results  \cite{CalabresePRL2006,calabrese2005evolution}, the light cone is  defined by the velocity of the fastest moving quasiparticles. It has also been shown that different types of excitations, including complex bound states, can be identified in the time evolution after a local quench~\cite{ganahl}. 
 
Here we investigate the non-equilibrium dynamics in a spin-1/2   XXZ chain after a local quench.
More specifically, we connect two chains in different phases -- one in the ferromagnet phase and the other in the critical  phase -- and investigate the dynamics via time-dependent density matrix renormalization group~\cite{dmrg3}. As usual for local quenches, we expect   that, in the thermodynamic limit and after sufficiently long times, the system will equilibrate to the ground state of the final Hamiltonian\cite{abraham1970thermalization}. In this case, the final state has nonzero magnetization inside the critical chain due to the proximity   with the ferromagnetic chain. Therefore, this can be viewed as a local quench to investigate how fast the order parameter of the ordered subsystem penetrates into the disordered one. We are particularly interested in the regime where the anisotropy parameter of the critical chain is close to the transition to the ferromagnetic phase.  In addition to the on-site magnetization, we   investigate the propagation of the bipartite entanglement entropy. We find  that the propagation of information in the
critical chain shows a light cone with the velocity of the fastest ``spinon'' excitations, which is known exactly from the Bethe ansatz solution of the XXZ model \cite{takahashi,korepin1997quantum}. More interestingly, when the anisotropy parameter approaches the ferromagnetic isotropic point, there are   bound state excitations~\cite{takahashi} which propagate faster than the spinons and create a second light cone with a greater velocity. The bound states we observe arise in the subspace of zero magnetization and as such are different from those investigated in Ref. \cite{ganahl}, which had smaller velocity than the spinons. For both spinon and bound state light cones, we find an  agreement  between the velocities calculated from our numerics and those given by Bethe ansatz,  demonstrating  that this property of the dispersion  of low-lying excitations manifests itself  in the out-of-equilibrium dynamics. Regarding the equilibration, we find that the relative distance of the local magnetization from its
equilibrium value decays faster with time and has smaller finite size effects for sites near the interface than in the bulk of the critical chain.

This paper is organized in the following way. In Sec. \ref{secII}, we present the   XXZ model and  its elementary  excitations. In Sec. \ref{secIII}, we discuss the quench protocol. The results are presented in Sec. \ref{secIV}; subsection \ref{secA} is devoted to our main results concerning the propagation velocities of entanglement entropy and magnetization, from which we observe the spinon and bound state light cones; in subsection \ref{secB} we discuss the asymptotic long-time behavior. Finally, Section \ref{secV} presents the conclusions.

\section{Model}\label{secII}

We consider the spin-1/2  XXZ chain with $N$ sites and open boundary conditions
\begin{equation}\label{XXZ}
 H=J\sum_{i\in I}  \left( S^x_i S^x_{i+1}+S^y_i S^y_{i+1} +\Delta S^z_i S^z_{i+1}\right),
\end{equation}
where $S^\alpha_i$, $\alpha=x,y,z$, are spin operators acting on site $i$ and $I$ denotes the
set of sites that compose each chain (see Sec. \ref{secIII} for more details). Here $J$ is the exchange coupling constant   and $\Delta$ is the anisotropy parameter. Throughout this paper we use $J=1$ as the unity of energy and set $\hbar=1$.

This model has exact solution by means of the Bethe ansatz \cite{takahashi,korepin1997quantum}.  The ground state phase diagram contains  three phases: a gapless, critical phase for $-1<\Delta\leq1$    and two long-range-ordered phases, a gapped N\'eel  phase for $\Delta>1$ and a ferromagnetic phase  for $\Delta\leq-1$.

The  XXZ model can be mapped, by a Jordan-Wigner transformation, into the spinless fermion model described by the   Hamiltonian \cite{giamarch}
\begin{eqnarray}\label{fermions}
 H&=&\sum_i \left [ - T ( c^{\dagger}_{i+1}c_i+c^{\dagger}_{i}c_{i+1}) \right. \nonumber \\
&&+ \left. V \left( c^{\dagger}_{i} c^{\phantom\dagger}_{i}-\frac12\right) \left( c^{\dagger}_{i+1} c^{\phantom\dagger}_{i+1}-\frac12\right)\right],
\end{eqnarray}
where $c_i$ are local fermionic operators, which satisfy the anticommutation relation $\{c^{\phantom\dagger}_l,c^{\dagger}_m\}=\delta_{l,m}$, $T=J/2$ is the hopping amplitude, and $V=J\Delta$ is the nearest-neighbor interaction strength. From these relations, $\Delta<0$ corresponds to an attractive interaction regime and $\Delta>0$ represents a repulsive interaction. For $\Delta=0$, we obtain  the XX model, which is equivalent to free spinless fermions. When analyzing our results, it may be helpful to think about spinless fermions instead of spins.


The Bethe ansatz solution of the XXZ model provides not only the ground state phase diagram, but also the full excitation spectrum. 
For $\Delta \leq -1$ the elementary excitations are called magnons and are gapless at the isotropic point  $\Delta=-1$ but have a gap given by $|\Delta|-1$ for $\Delta<-1$. The magnon dispersion relation is
\begin{equation}
 E_m(k)=-J(\Delta +\cos k), \label{eqmagnon}
\end{equation}
from which we obtain the maximum magnon velocity $v_m= \text{max}\{\left|\frac{dE_m}{dk}\right|\}=J$.

In the critical  phase, the  elementary excitations, known as spinons, correspond to single holes in the ground state root density \cite{takahashi,korepin1997quantum}.  Their exact dispersion relation is given by \be
\epsilon_s(k)=v_s\sin k,\qquad (0<k<\pi)
\ee
where
\begin{equation}\label{vel}
 v_s=\frac{\pi \sqrt{1-\Delta^2}}{2\arccos \Delta}
 \end{equation}
can be identified with the maximum value of the spinon velocity:\be
 \text{max}\left\{\frac{d\epsilon_s}{dk}\right\}=v_s.
 \ee

\begin{figure}[t]
 \centering
 \includegraphics[width=.7\columnwidth]{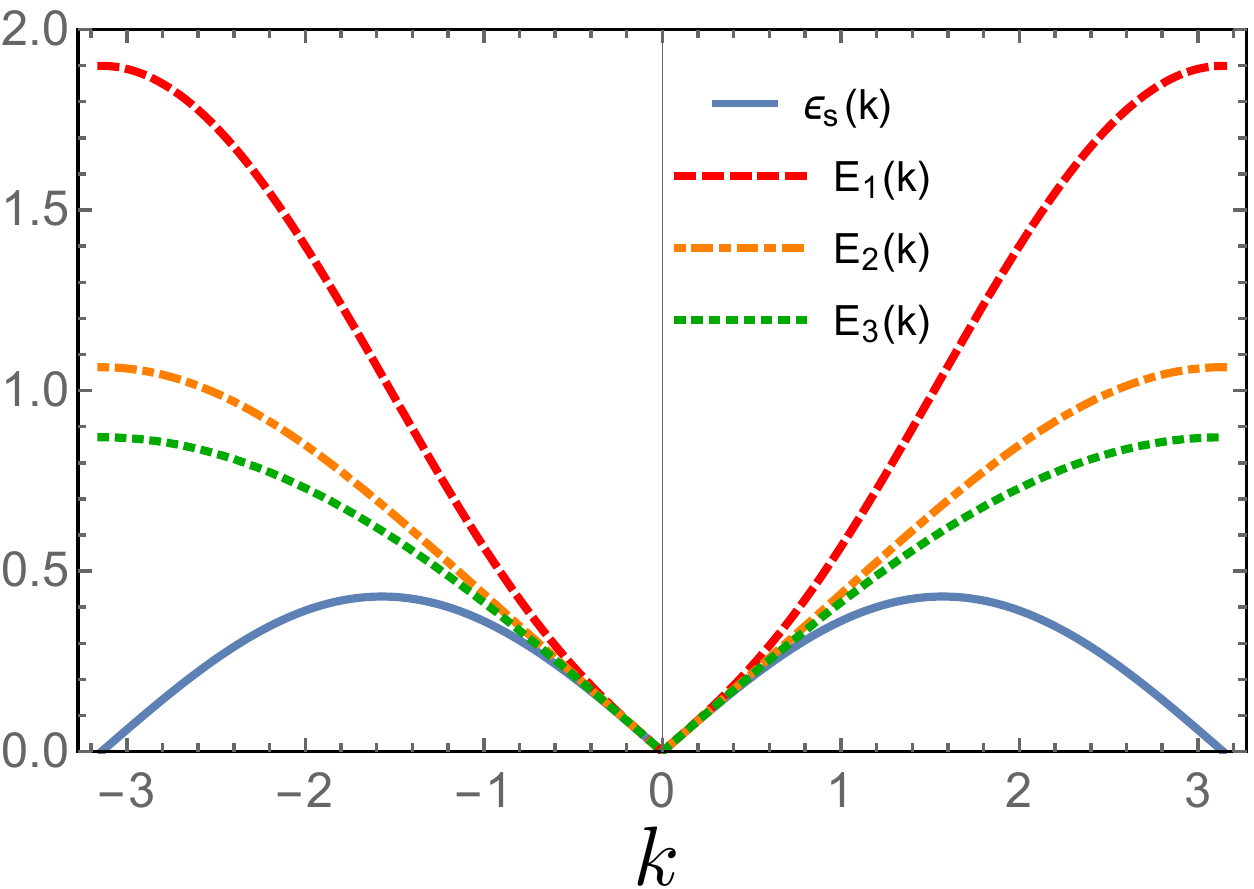}
 \caption{(Color online) Exact dispersion relation for spinons $\epsilon_s(k)$ and for the three bound state branches $E_n(k)$ for $\Delta=-0.75$. The bound state velocity $v_b\approx0.866$ is given by the slope of $E_1(k)$ at the inflection point $k_0\approx1.533$. }
 \label{dispersion}
 \end{figure}

The Bethe ansatz also  allows for low-lying excited states with complex rapidities, called strings \cite{takahashi}, which can be interpreted as bound states of the elementary particles. In the  subspace of zero magnetization, bound states form above the two-spinon continuum for $-1<\Delta<0$. The dispersion relation for the length-$n$ string is given by \cite{takahashi} \begin{eqnarray}\label{disp}
 E_n(k)&=&\frac{\pi\sqrt{1-\Delta^2}}{\arccos \Delta}\left|\sin\left(\frac{k}{2}\right)\right|\times \nonumber \\
 &&\times\sqrt{1+\cot^2\left[\frac{n\pi}{2}\left(\frac{\pi}{\arccos \Delta}-1\right)\right]\sin^2\left(\frac{k}{2}\right)},\nonumber\\
\end{eqnarray}
where $n=1,2,...,\left\lfloor\frac{\arccos \Delta}{\pi-\arccos \Delta}\right\rfloor$ and $\lfloor x \rfloor$ is the floor function. Note that the number of bound state branches depends on $\Delta$, but the $n=1$ branch exists for any $\Delta<0$. In the fermionic picture, this simplest bound state   can be viewed as being composed by a particle and a hole that interact with each other and bind for arbitrarily weak attractive interactions  \cite{rodrigo1}, in close analogy with  the formation of Wannier excitons in one dimension; in the corresponding spin scenario (particle represents $\uparrow$ spin and hole, $\downarrow$ spin), this bound state is an excitation with zero magnetization.
For $\Delta\to-1$, the $n=1$ bound state approaches the dispersion of magnons at the ferromagnetic isotropic point:\be
\lim_{\Delta\to-1}E_1(k)=1-\cos k.
\ee
The spinon and bound state dispersion relations for $\Delta=-0.75$ are illustrated in Figure \ref{dispersion}.

Equation (\ref{disp}) predicts that the maximum velocity of bound state excitations is obtained  for $n=1$ and is given by \be
v_b=\text{max}\left\{v(k)\right\},\label{velb}
\ee where $v(k)=dE_1(k)/dk$.
One can check that $k=0$ is a stationary point of $v(k)$ for $-1<\Delta<0$. For small $|\Delta|$, $v(k)$ is a concave function; in this case, the maximum bound state velocity occurs at $k=0$ and coincides with $v_s$. On the other hand, $k=0$ is a local minimum if $\Delta$ is close to $-1$. Therefore, there is a $\Delta^*$ such that $k=0$ is an inflection point for $v(k)$: 
\be
\left.\frac{d^3E_1}{dk^3}\right|_{k=0,\Delta=\Delta^*}=0.
\ee
The solution is \be
\Delta^*=\cos\left(\frac{3\pi}{5}\right)=\frac{1-\sqrt{5}}{4}\approx -0.309.
\ee
For $-1<\Delta<\Delta^*$, the maximum bound state velocity occurs at $k=k_{0}>0$ given by the inflection point of the bound state dispersion (see Figure \ref{dispersion} for an example). As a result, we obtain $v_b>v_s$, which means that $n=1$ bound states can propagate faster than spinons. Note that in this cases $v_b$ is \emph{not} a low-energy property, but depends on the bound state dispersion at finite energies.





\section{Local Quench}\label{secIII}

\begin{figure}[t]
 \centering
 \includegraphics[width=\linewidth]{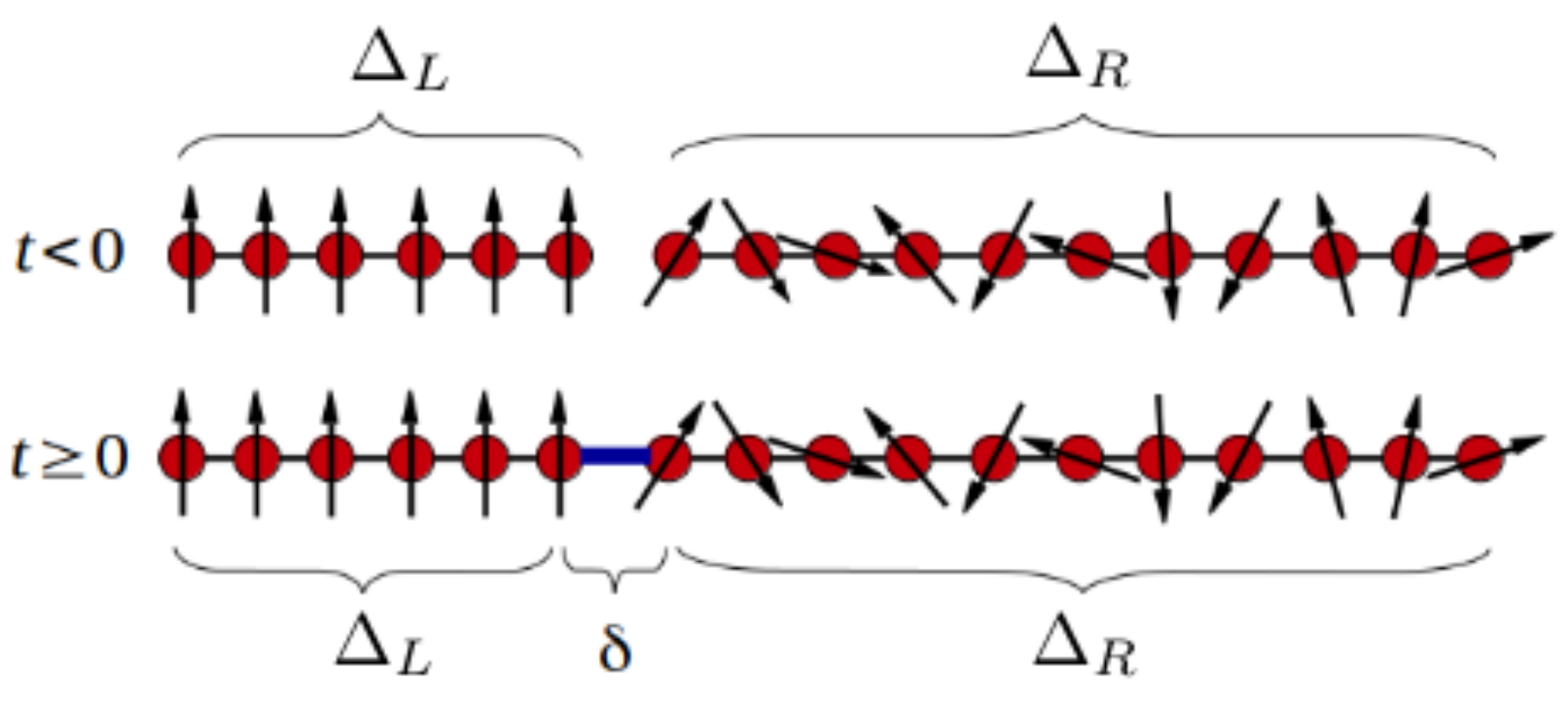}
 \caption{(Color online) Representation of our quench protocol: for $t<0$ the chains are prepared in their respective ground states; at $t=0$ they are connected. The left chain is in the ferromagnetic phase and the right one is in critical phase.}
 \label{cadeia}
 \end{figure}

We consider the following quench protocol (see Figure \ref{cadeia}): two finite chains with different $\Delta$ are initially separated and prepared in the  ground state of their respective Hamiltonians, $|G_L\rangle$ and $|G_R\rangle$ for the left and right chains. At time $t=0$, the chains are connected and we let the system evolve. The Hamiltonian of the whole system is given by
\begin{equation}\label{hamiltonian}
 H(t)=H_L +\Theta(t) \left( S^x_k S^x_{k+1}+S^y_k S^y_{k+1} + \delta S^z_k S^z_{k+1}\right) + H_R,
\end{equation}
where $\Theta(t)$ is the Heaviside step function, 
and $H_{L}$ and $H_{R}$ are the Hamiltonians of the left and right chains, respectively. These are described by Eq. (\ref{XXZ}), with
$I=\{1,2,...,k-1\}$ for the former and $I=\{k+1,k+2,...,N-1\}$ for the latter. The index $k$ thus labels the last site of the left chain. After the quench, the   exchange coupling at the junction between chains  is set to   $J=1$, while the anisotropy parameter  $\delta$ can assume any constant value. In this case, we have three free parameters: $\Delta_{L(R)}$, which defines the phase of the left (right) chain, and $\delta$, which sets the coupling strength between the chains.
Note that the simple fact of connecting the two chains is sufficient to create the non-equilibrium dynamics, since the initial state, which is the product of the ground states of the separate chains, is {\it not} an eigenstate of the new Hamiltonian.

Works in the literature \cite{cut1,cut3,cut4,cut6} have analyzed the  growth of entanglement across  the junction  after connecting chains in the same phase. Here, we connect chains in different phases, with the goal of investigating how one chain affects the properties of the other. More specifically, we study the changes produced by the quench over
the magnetization per site
\begin{equation}
\langle S_i^z(t) \rangle=\langle\Psi(t)|S_i^z|\Psi(t)\rangle \label{magpersite}
\end{equation}
and the bipartite entanglement entropy
\begin{equation}\label{entropy}
S(x,t)=-\sum_i\lambda_i(x,t)\ln\lambda_i(x,t),
\end{equation}
where $\lambda_i(x,t)$ is the eigenvalue of the reduced density matrix, $\rho(x,t)$, associated with the   partition $1\leq j\leq x$ at time $t$.


In particular, we maintain the left chain in the ferromagnetic phase, $\Delta_L \leq-1$, and the right chain in the critical phase,  $-1 < \Delta_R  \leq 1$.  In addition, in the left chain, we apply a very small magnetic field in the first site, which breaks the degenerescence in the spin orientation and selects the $\uparrow$ spin state. In this way, the ground state of the left chain presents a well defined magnetization, with all spins aligned in the same direction, while the ground state of the right chain, $|G_R(\Delta_R)\rangle$, shows no magnetic order. The initial state of the system, for this choice of parameters, can be written as
\begin{equation}\label{estado}
|\Psi(0)\rangle=|\uparrow\uparrow\uparrow\uparrow \ldots\uparrow\uparrow\rangle \otimes |G_R(\Delta_R)\rangle.
\end{equation}

As mentioned in the previous section, the XXZ Hamiltonian without external magnetic field is mapped, through Jordan-Wigner transformation, into a spinless fermion model with chemical potential $\mu=-V/2$ [see Eq. (2)], which ensures particle-hole symmetry. As particles correspond to $\uparrow$ spins and holes represent $\downarrow$ spins, particle-hole symmetry is equivalent to zero magnetization in the spin scenario. This is the case for the initial state of the right chain.
After the quench, only the total magnetization of the system (i.e. of the connected chains) is conserved. Nonetheless, we observe that, far from the junction, the right chain relaxes to a state in which the local magnetization is close to zero. Moreover, the relaxation dynamics can be approximately described by elementary excitations on top of the ground state with $S^z=0$ (see next section).

Figure \ref{cadeia} shows the quench protocol, as well as the real configuration of the ground state of the left chain before the quench. In most part of our results, the entire connected chain  has $N=80$ sites. We choose $k=N/4$, namely, the left chain has $N_L=20$ sites and the right one has $N_R=60$ sites.

%

\section{Results}\label{secIV}

In this section, we   discuss the effects of the quench on the magnetization and the entanglement entropy dynamics of the chains.
All the results for the post-quench dynamics were obtained via time-dependent DMRG calculation with a second order Suzuki-Trotter decomposition. We use a time steep $dt=0.05$, which keeps the error of the order of $10^{-8}$ for the time interval we consider.

\subsection{Light-cone effect}\label{secA}

\begin{figure}[b]
 \centering
 \includegraphics[width=\linewidth]{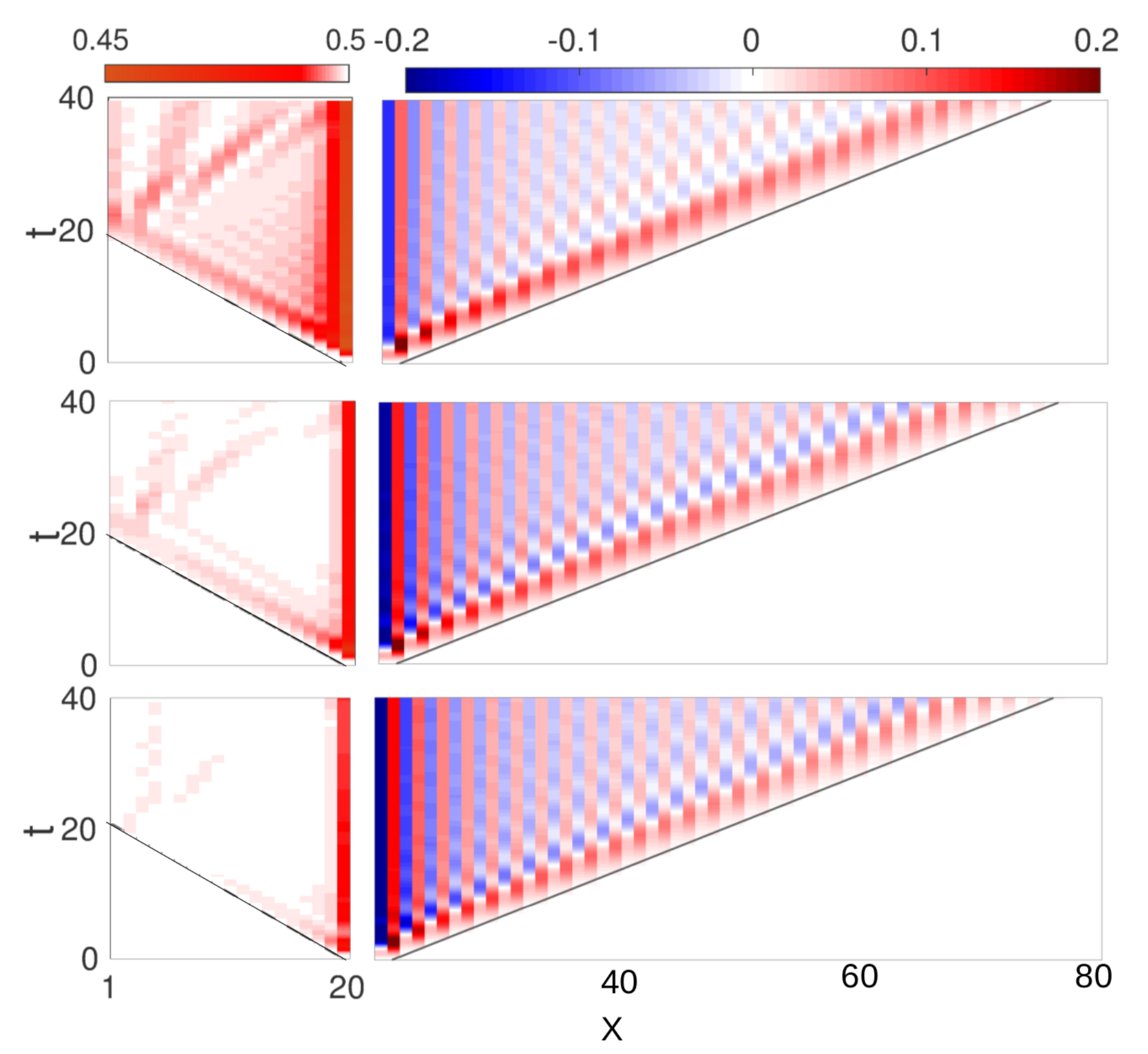}
 \caption{(Color online) Color map of the on-site magnetization as a function of time, $t$, and site, $x$, in both left and right chains (left and right columns, respectively). We decrease the value of the anisotropy parameter at the left chain, $\Delta_L=\{-1.1, -1.5,-2.0\}$ from top to bottom, while fix the value of $\Delta_R=0.5$ and $\delta=\Delta_R$. One can see the formation of light cones on the left and right chains, bounded by the maximum magnon and spinon velocities, respectively. }
 \label{Fig}
 \end{figure}

Connecting  the two chains drives the system away from equilibrium, in such a way that the information about the change in the Hamiltonian starts propagating from the interface between the chains.
After the quench, as the system evolves, changes in magnetization and entanglement entropy flow over the chains, forming effective light cones, as we explore in this subsection.
An example can be seen in Figure~\ref{Fig}, which shows the on-site magnetization profile for the connected chain as a function of time and site for fixed $\Delta_R=0.5$ and three different values of $\Delta_L=\{-1.1, -1.5,-2.0\}$; we can see the formation of light cones in both left and right chains, with corresponding different velocities.

When we decrease $\Delta_L$ of the left chain (left column of the figure, from top to bottom), we see that the amplitude of the on-site magnetization decreases, while the velocity that bounds the light cone remains the same and is given by the maximum magnon velocity, $v_m=J=1$.     
These results indicate that, for $\Delta_L$ close to the isotropic ferromagnet point, the quench produces a perturbation in the on-site magnetization that propagates with $v_m=J$ independent of the value of $\Delta_L$, in accordance with expected for the velocity obtained from Eq. (\ref{eqmagnon}). For the parameters considered in the figure, the difference $\Delta S(x,t)=S(x,t)-S(x,0)$ of entanglement entropy in the left chain also defines a light cone bounded by $v \approx J$, with amplitude that becomes smaller as we decrease $\Delta_L$ (not shown).

The perturbation observed in the left chain is small, since it is in a gapped phase. As we decrease $\Delta_L$, the ferromagnetic gap increases and the state of the left chain becomes more insensitive to the quench; for $\Delta_L \ll -1$ the perturbation created by the local quench does not penetrate far into the ferromagnet. In fact, in the ground state of the final Hamiltonian (Eq. \ref{hamiltonian}), to which the system equilibrates after long times, the spins on the left chain are close to being fully polarized. In the fermionic picture, this regime of large negative $\Delta_L$ corresponds to a strong attractive interaction. As a result, the occupation of each site of the left chain by a fermion prevents a significant change in its fermionic density.

The velocity corresponding to the light cone seen in the right chain (right column of Figure \ref{Fig}) depends only on the parameters of this chain, being independent of $\Delta_L$. Our results indicate that for $\Delta^*\lesssim \Delta_R <1 $ the light cone  is defined by the maximum spinon   velocity  $v_s$ (hereafter called the spinon  light cone).
On the other hand, for $-1<\Delta_R \lesssim \Delta^*$, a second wavefront appears in front of that corresponding to spinons, meaning that changes in magnetization and entanglement can propagate faster than $v_s$ due to the presence of bound state excitations. In this parameter regime, we observe, for both magnetization and entanglement, a second light cone (called the bound state light cone) outside the spinon one.

\begin{figure}[b]
 \centering
 \includegraphics[width=\linewidth]{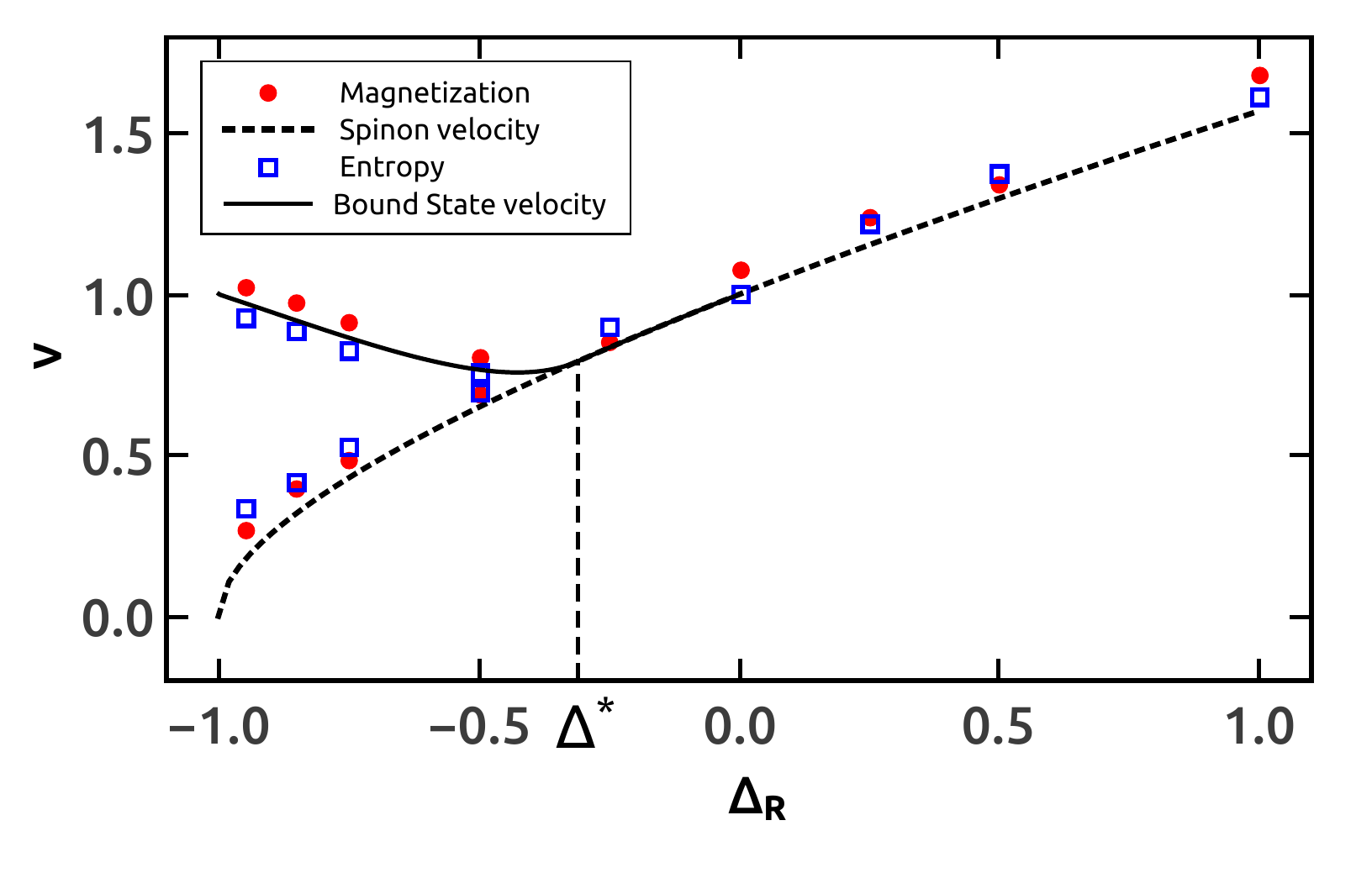}
 \caption{(Color online) Wavefront propagation velocity of magnetization (circles) and entropy (squares). The dashed line is the maximum spinon excitation velocity, given by Eq.~(\ref{vel}), and the solid curve is the maximum velocity in the case of bound state excitations, which can be obtained from Eqs.~(\ref{disp}) and (\ref{velb}).}
 \label{velocidade}
 \end{figure}

 Our main results are summarized in Figure~\ref{velocidade}. Circles and squares are, respectively, magnetization and entanglement wavefront velocities obtained from the light cones seen in the right chain in our simulations.
The dotted curve corresponds to the maximum spinon velocity, as predicted by Bethe ansatz [Eq. (\ref{vel})], while the solid line gives the maximum bound state velocity, calculated from Eqs. (\ref{disp}) and (\ref{velb}). 
We can see that for $\Delta_R<\Delta^*$ the maximum bound state   velocity becomes larger than the spinon velocity.  The velocities obtained from our numerics agree well with either the spinon or the bound state excitation velocity. This demonstrates that the velocity of excitations can be detected  in the dynamics after the local quench. Both observables, on-site magnetization and bipartite entanglement entropy, can be used to detect the velocities, giving very similar results.

The rich dynamics observed in the right chain do not depend on the size of the left chain or on the value of $\Delta_L$. In the following, for simplicity, we fix the left chain deeply in the ferromagnetic phase, with $\Delta_L=-20$, and analyze in detail the light cones seen in the right chain.

 
 \subsubsection{Spinon light cone}

\begin{figure}[b]
 \centering
\includegraphics[width=\linewidth]{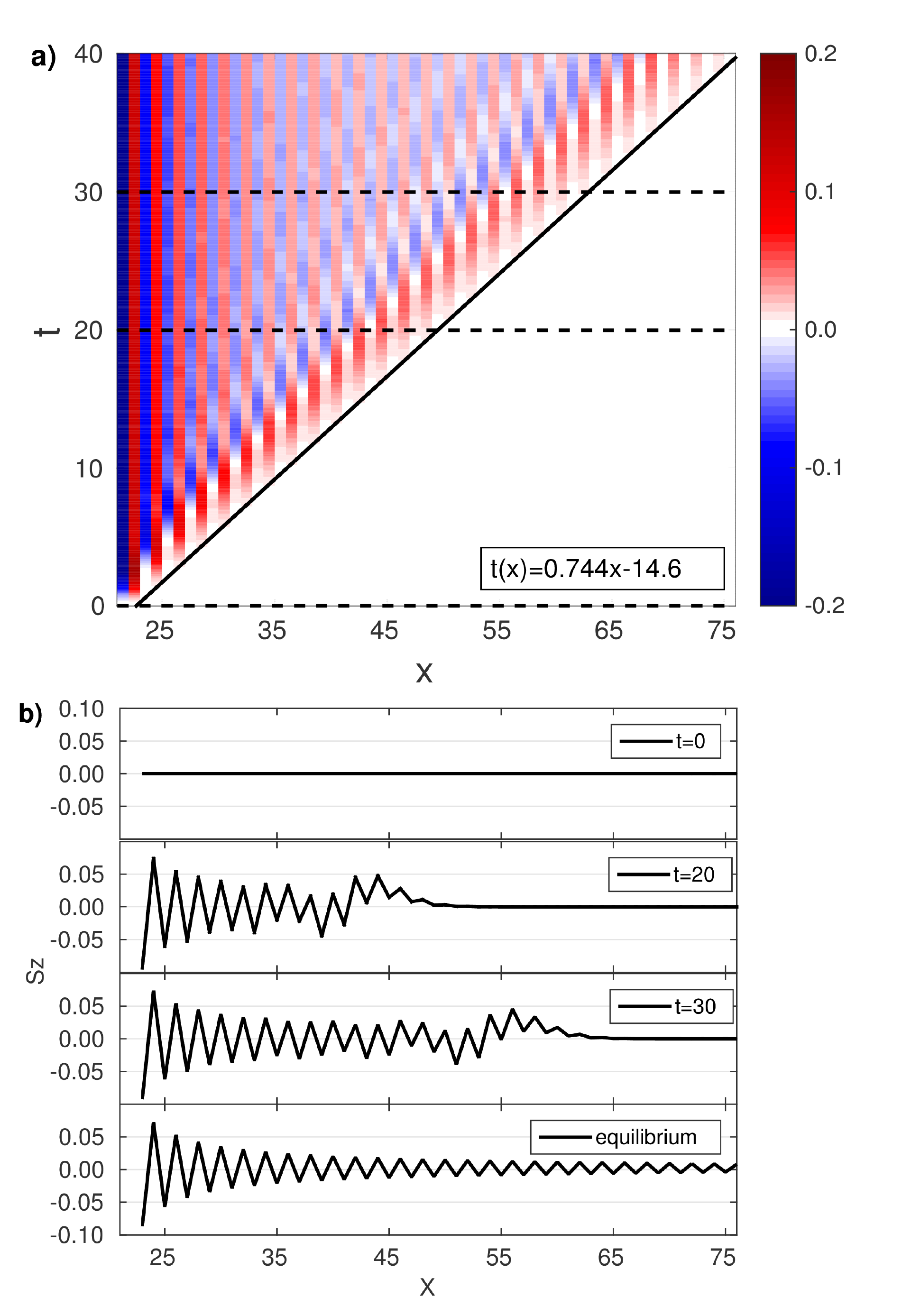}
 \caption{(Color online) a) Color map of magnetization in the right chain when $\Delta_R=0.5$, $\Delta_L=-20$, and $\delta=\Delta_R$. The dotted lines indicate cuts in time, while the solid curve corresponds to a linear fit to the wavefront. b) Spin profile in the right chain for the times indicated in panel a). The last plot corresponds to the equilibrium state.}\label{cortes_05}
\end{figure}

 Figure \ref{cortes_05} shows the dynamics of the on-site magnetization in the right chain for $\Delta_R=0.5$.
At time $t=0$, the magnetization vanishes, as expected for the ground state of the critical phase. After the chains are connected,
the left chain induces some magnetization in the right chain. 
Effectively, the left chain plays the role of a boundary magnetic field along the $z$-direction acting on the first site of the right chain, whereas $\delta$ defines the coupling to this field.
The amplitude of
$\delta$ does not interfere in the wavefront velocity, which is a property of the bulk excitations; for simplicity we set $\delta=\Delta_R$. The sign of $\delta$ determines a small polarization of the first spin of the right chain: if $\delta>0$ the interaction is antiferromagnetic, making the site to anti-align with the last site of the left chain; 
instead, if $\delta<0$ the coupling is ferromagnetic, forcing them to align. Once the first site of the right chain defines its magnetization, it induces some magnetization in the next site and this effect continues throughout the chain. This change in magnetization propagates like a wave, with a well-defined velocity that depends only on the parameters of the right chain. The same behavior is observed for other values of $\Delta_R>0$ in the critical  phase. 

Figure \ref{cortes_05} a) shows a color map of the magnetization as a function of time and position. From this we clearly see that the magnetization flow is bounded, defining an effective light cone. The solid black line corresponds to a linear fit to the wavefront, which is defined at the time at which the magnetization reaches $0.01\langle S_{max}^z\rangle$, where $\langle S_{max}^z\rangle$ is the maximum amplitude of the magnetization assumed by the spins in the right chain; circles in Figure~\ref{velocidade} correspond to the inverse of this line slope. According to the comparison shown in Figure~\ref{velocidade}, the propagation velocity corresponds to the maximum spinon   velocity.

The wavefront has a positive magnetization. However, inside the light cone the chain shows some staggered magnetization. This effect is expected since at long times the system must approach the equilibrium state, in which the response of the critical chain to a boundary magnetic field shows Friedel-type oscillations \cite{EggerPRL1995,FabrizioPRB1995,KitanineJSTAT2008}.
 
 
In panel b) of Figure \ref{cortes_05} we show constant-time cuts of  the spin profile in the right chain   [corresponding to the dotted lines in panel a)].  Note that  inside the light cone the spins stay permanently polarized. The last plot illustrates the spin profile for the chain in the equilibrium state, which corresponds to the ground state of   $H(t>0)$ in Eq.  (\ref{hamiltonian}). We discuss the approach to this equilibrium state in the next subsection.

\begin{figure}[t]
 \centering
 \includegraphics[width=\linewidth]{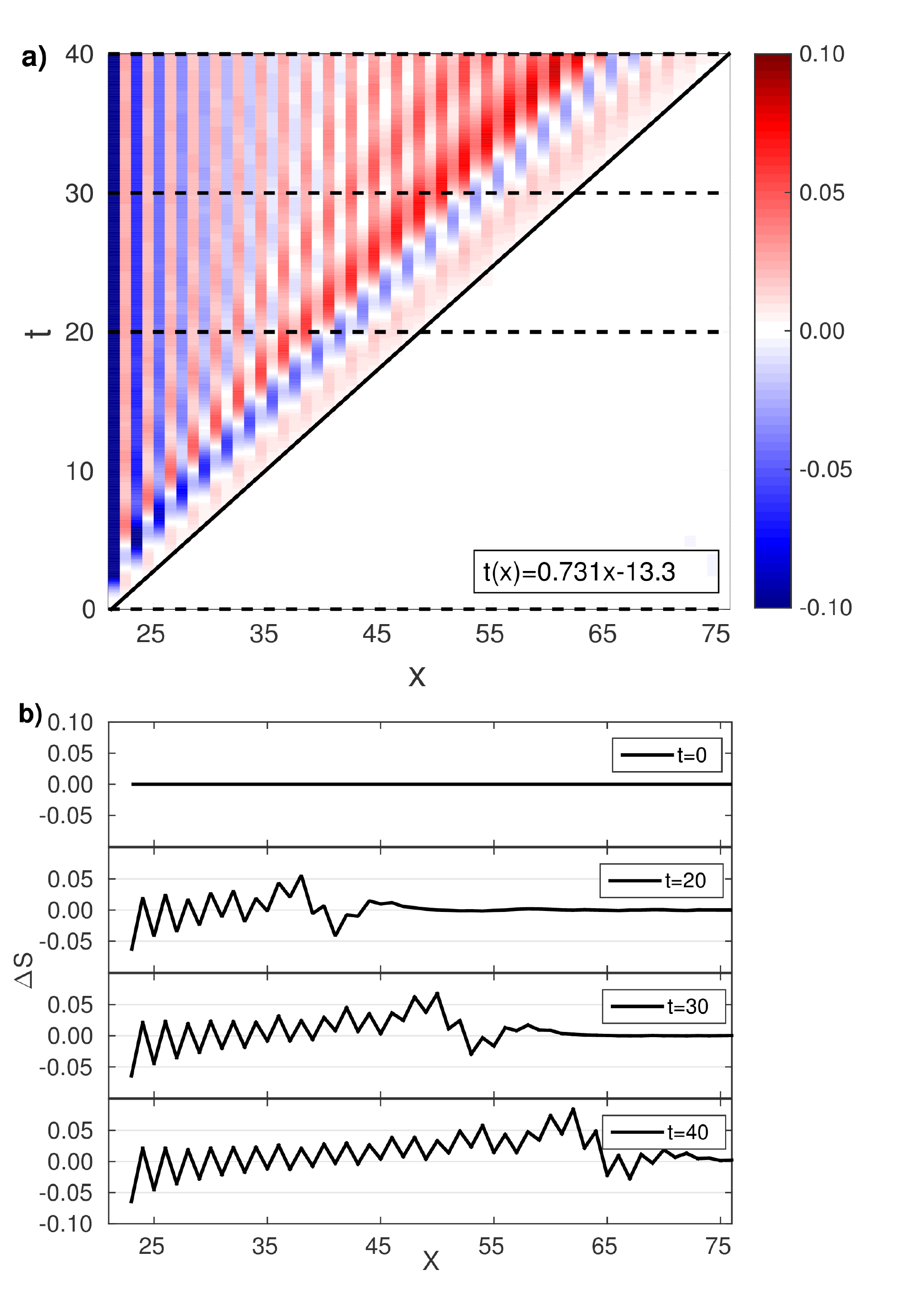}
 \caption{(Color online) a) Color map of the difference of entropy in the right chain when $\Delta_R=0.5$, $\Delta_L=-20$, and $\delta=\Delta_R$. The dotted lines indicate cuts at the times considered in panel b), while the solid line is a linear fit to the wavefront. b) $\Delta S$ profile in the right chain at $t=0$, $t=20$, $t=30$, and $t=40$.}
 \label{collor_map_entro}
 \end{figure}

Now, let us investigate the behavior of the entropy [defined in Eq. (\ref{entropy})] for the same parameter regime $\Delta>\Delta^*$. Since the ground state of the critical phase has a finite entropy~\cite{entropia}, we analyze the difference $\Delta S(x,t)=S(x,t)-S(x,0)$ of entropy created by the quench, which we present in Figure~\ref{collor_map_entro}.
Panel a) shows the color map of $\Delta S(x,t)$. Again we observe a  light cone as in the case for the magnetization. Indeed, it can be demonstrated analytically that the entropy dynamics must obey a light-cone effect \cite{raphael}. Interestingly, magnetization and entropy wavefronts travel with the spinon excitation velocity, as can be concluded from Figure~\ref{velocidade} (squares in this figure correspond to linear fits to the entropy wavefronts, as the one seen in Figure~\ref{collor_map_entro}).

Figure~\ref{collor_map_entro} b) presents $\Delta S(x,t)$ profiles at certain instants of time, corresponding to the dotted lines in the color map, from which we can follow the propagation of the disturbance caused by the quench along the right chain.
Close to the wavefront there is an increase in the entropy; after the disturbance passes, the difference in entropy becomes negative.  Note that the amplitude of the difference of entropy is rather small, of the order of $10^{-1}$. Inside the light cone, the difference of entropy presents site dependent oscillations, similarly to those in the magnetization data. However,
in contrast with the Friedel oscillations in the magnetization, the oscillations in the entanglement entropy are  present already in the initial state of the open chain~\cite{entropia2}. The effect of the quench is to decrease the  amplitude of these oscillations (not shown).

We have observed the spinon light cone for other values of $\Delta$ in the critical phase, from which we have obtained the circles and squares shown in Figure~\ref{velocidade}. Interestingly, for $\Delta<\Delta^*$, we can define a second light cone, which we analyze in detail below.

\subsubsection{Bound state excitation light cone}


\begin{figure}[t]
 \centering
  \includegraphics[scale=0.45]{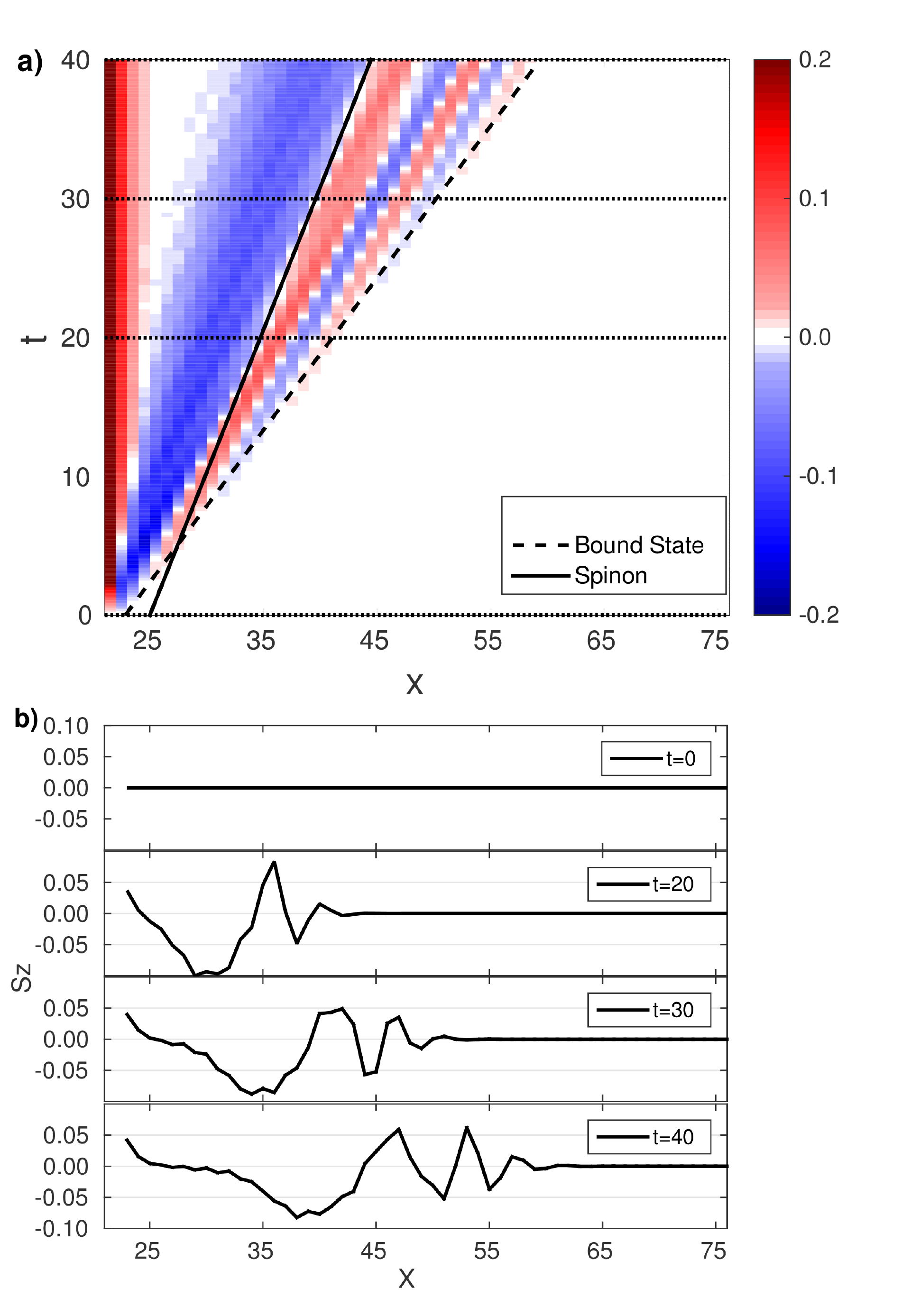}
  \caption{(Color online) a) Time evolution of the magnetization in the right chain when $\Delta_R=-0.75$, $\Delta_L=-20$, and $\delta=\Delta_R$. The horizontal dotted lines indicate cuts at certain times, while the solid and the dashed curves delimit the spinon and bound state excitation light cones.
 b) Spin profiles in the right chain at the times indicated by dotted lines in panel a).}
  \label{collor_map_-075}
\end{figure}

Figure~\ref{collor_map_-075} a) shows the magnetization in the right chain as a function of time and site index for $\Delta_R=-0.75$; panel b) of the figure shows cuts at fixed times. 
There are two main distinctive features: first,  the wavefront followed by a region with oscillating magnetization; second, a large pulse with negative magnetization lagging behind the first region. The front of the  large pulse propagates approximately with the maximum spinon  velocity $v_s$. The positions of this front as a function of time thus define the spinon  light cone, indicated by the solid line in panel a). To obtain the propagation velocity presented in Figure~\ref{velocidade} we consider only long times; this way we ensure that the large pulse is formed.
Furthermore, by analyzing the front of the oscillations that form before the negative large pulse, we conclude that its positions   as a function of time can also be fitted to a straight line, defining a second light cone (see the dashed line in the color map of Figure~\ref{collor_map_-075}). Interestingly, the associated velocity [illustrated by a circle in Figure \ref{velocidade}] corresponds to the maximum velocity $v_b$ of bound state excitations, suggesting that these oscillations are signatures of bound states. Note that, although we can only measure the length-$1$ string velocity [which is related to $n=1$ in Eq. (7)], since it corresponds to the maximum velocity and as such is the one that defines the light cone, other bound states can be present in the system.

\begin{figure}[t]
 \centering
  \includegraphics[scale=0.45]{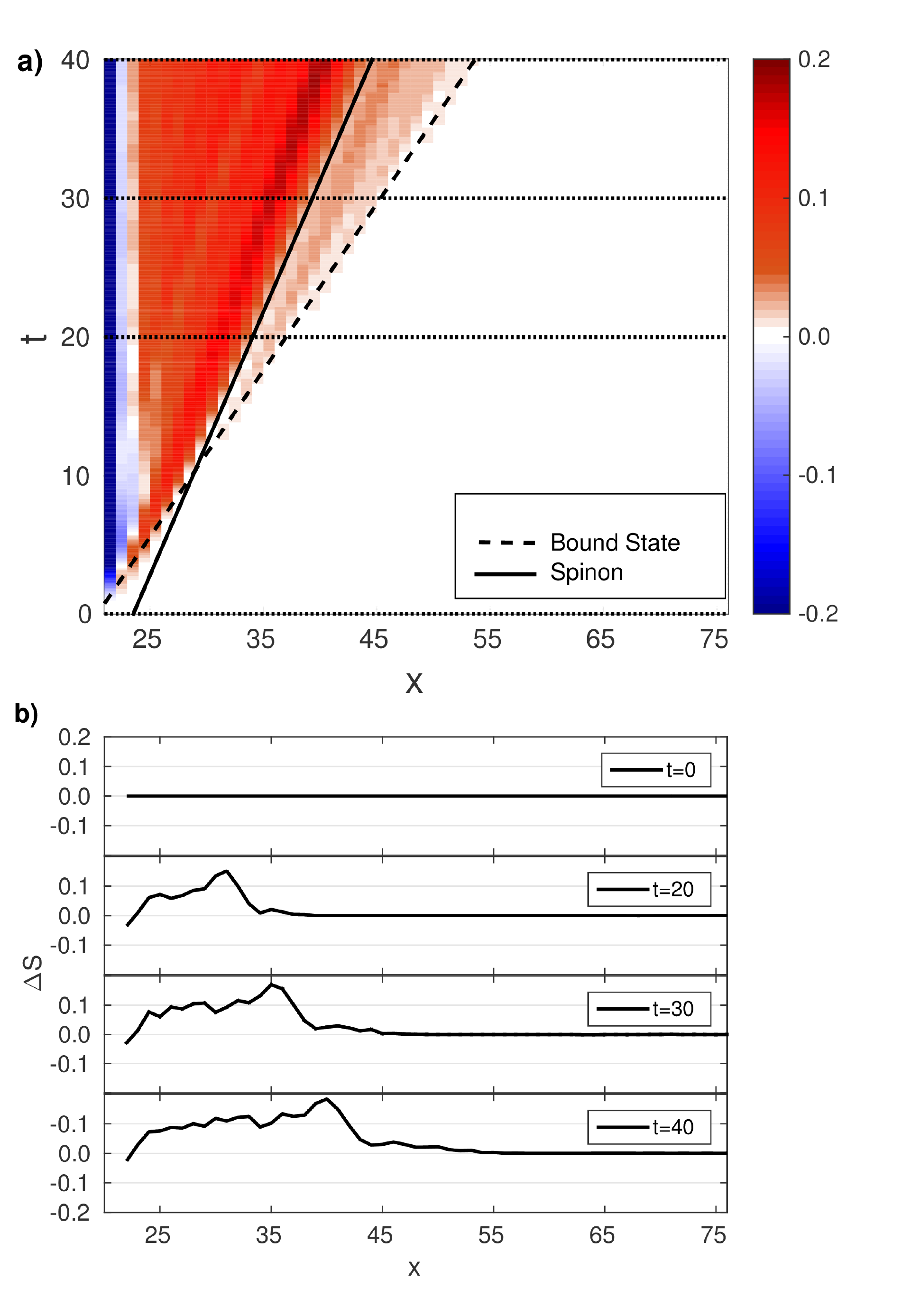}
  \caption{(Color online) a) Time evolution of $\Delta S(x,t)$ in the right chain when $\Delta_R=-0.75$, $\Delta_L=-20$, and $\delta=\Delta_R$. The horizontal dotted lines indicate cuts at times considered in panel b). The solid curve signals the spinon   light cone while the dashed one indicates the bound state   light cone.
 b) $\Delta S(x,t)$ profiles in the right chain at different times.}
  \label{collor_map_entro-075}
\end{figure}

Figure \ref{collor_map_entro-075} shows the  difference in entropy with respect to the initial state, $\Delta S(x,t)$. In panel a) we present the color map and in b) we show cuts at the times indicated by dotted lines in panel a).
Similarly to the propagation of the magnetization, one can recognize two different regions: at intermediate $x$ there is a large increase in the entropy, while at larger $x$ there is only a small increase in it. According to our analysis, the front of the former region propagates with the velocity $v_s$, whereas that of the latter propagates with velocity $v_b$.  In fact, in panel a), one can clearly identify two light cones, one related to spinons  (solid curve) and the other to bound states (dashed line). As observed for the magnetization, the propagation velocities obtained from the entropy difference, represented by square points in Figure \ref{velocidade}, are in  good agreement with the analytical results.


\begin{figure}[t]
 \centering
  \includegraphics[scale=0.24]{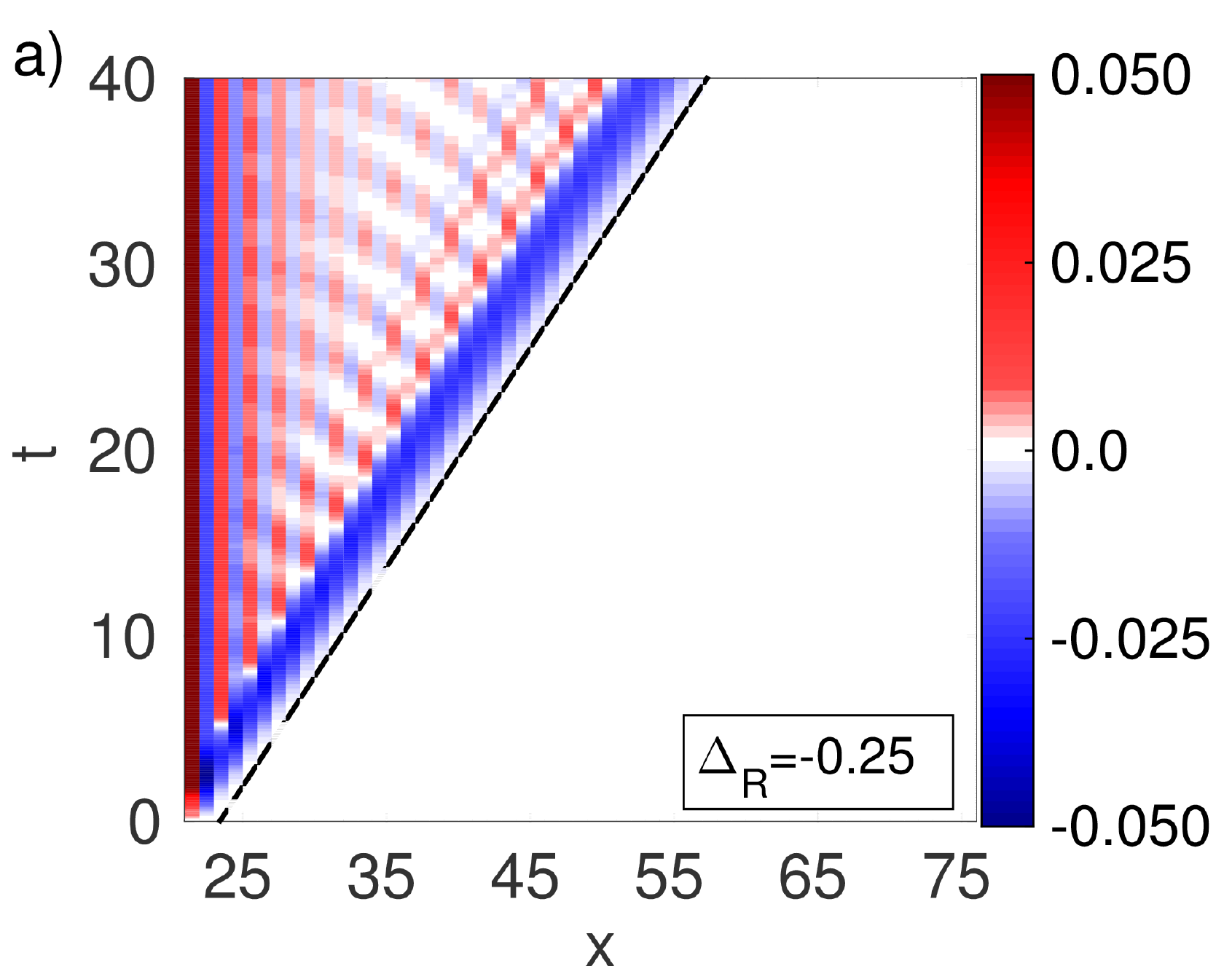}
  \includegraphics[scale=0.24]{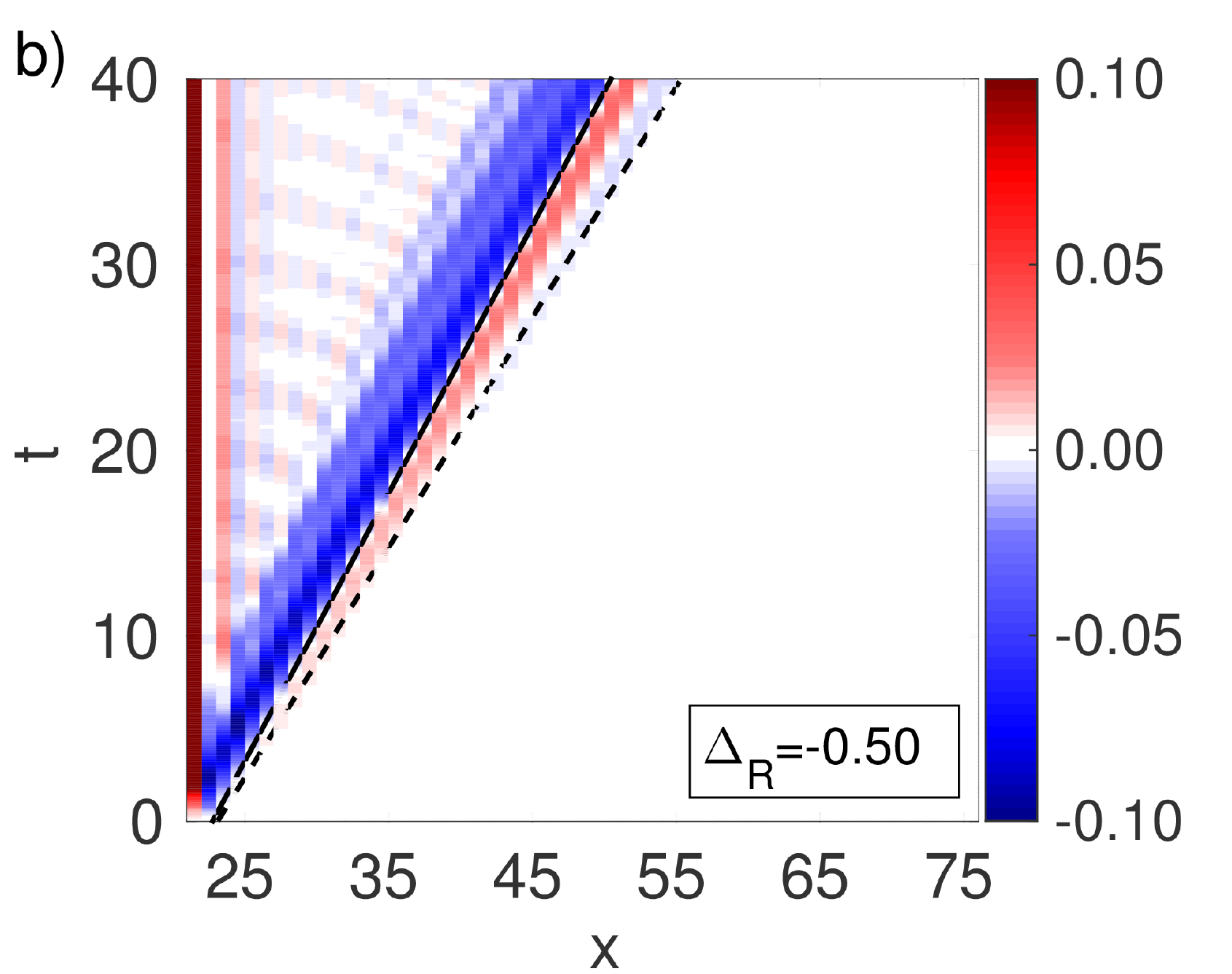}
  \includegraphics[scale=0.24]{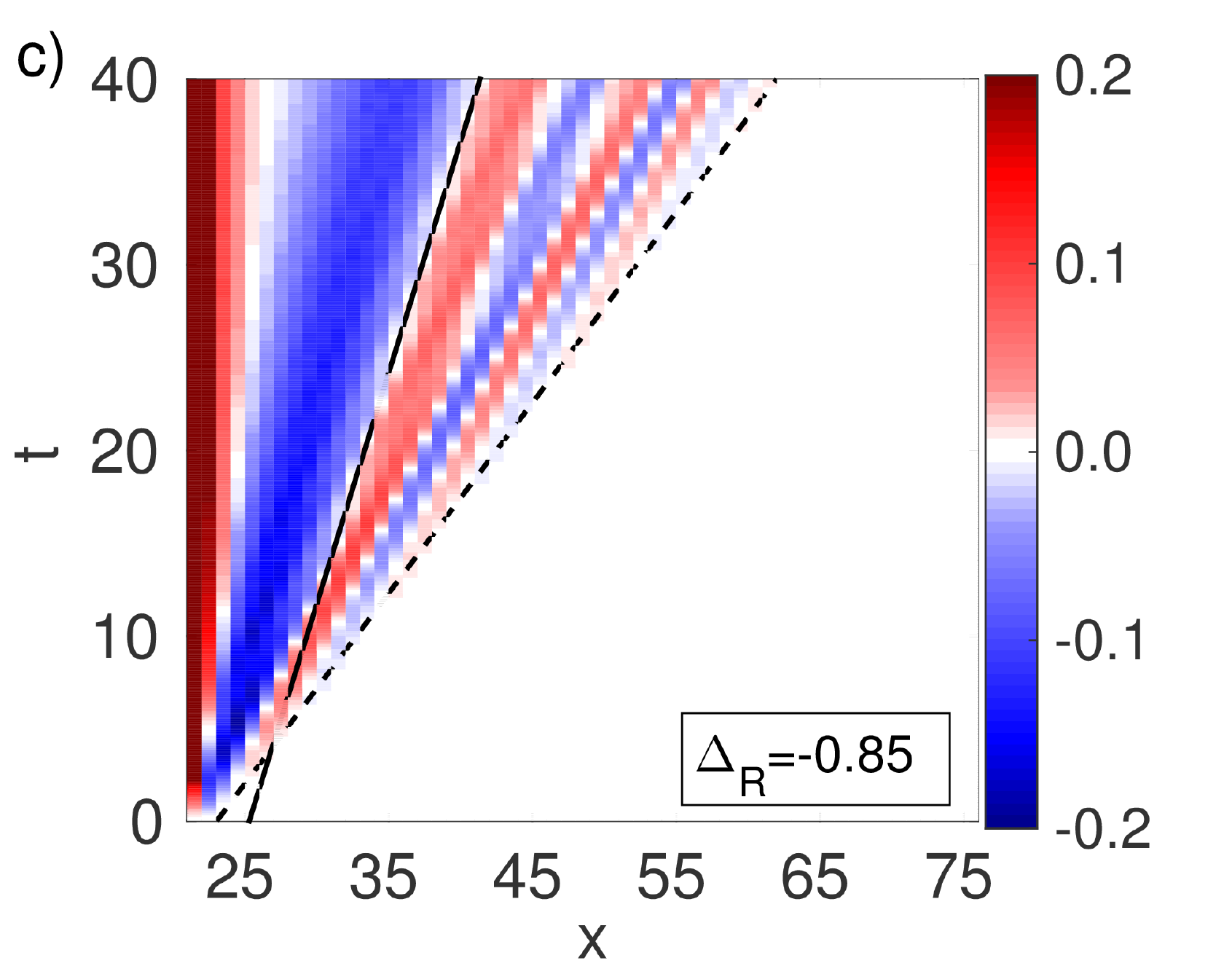}
  \includegraphics[scale=0.24]{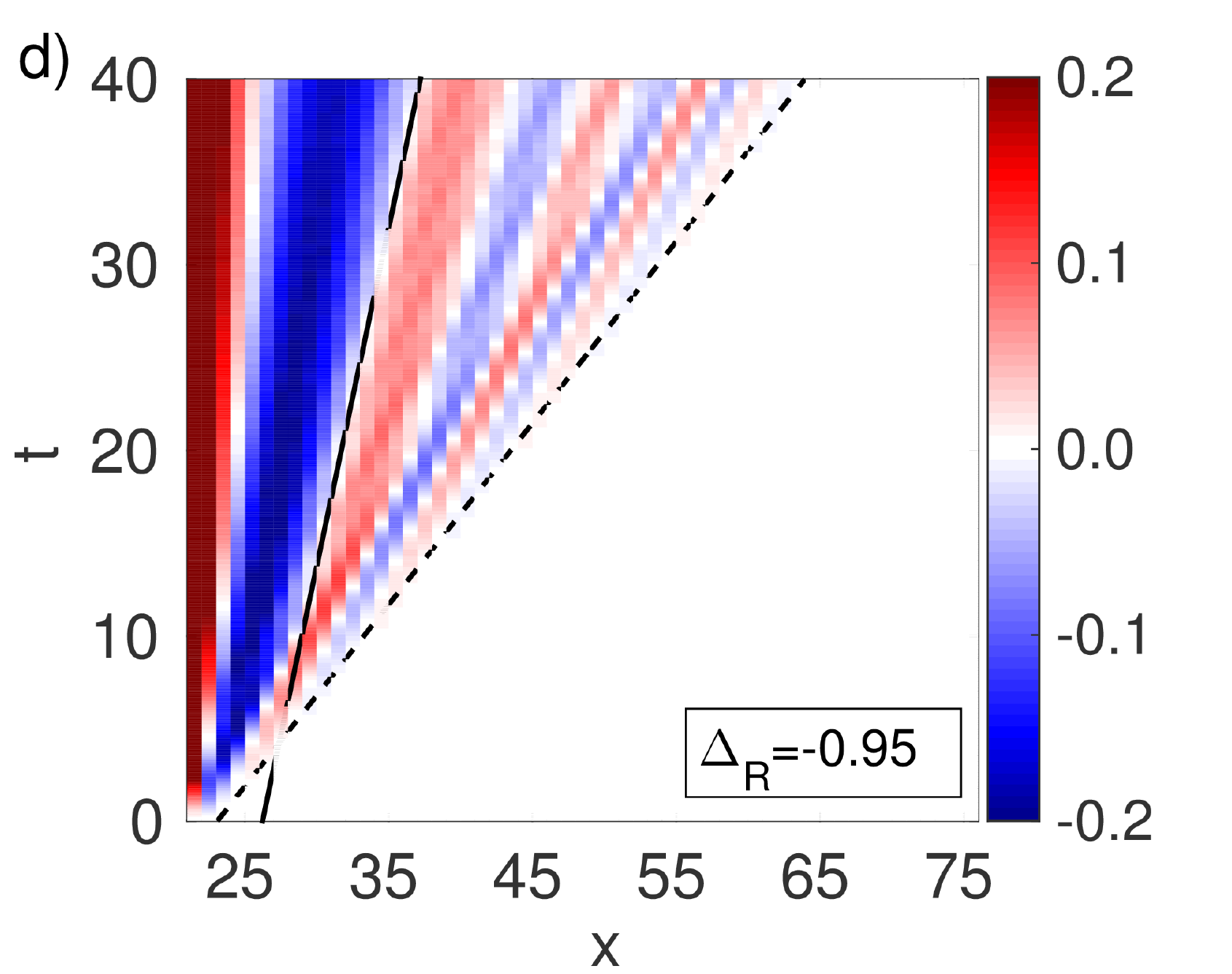}
  \caption{(Color online) Color map of magnetization in the right chain when a) $\Delta_R=-0.25$, b) $\Delta_R=-0.50$, c) $\Delta_R=-0.85$, and d)~$\Delta_R=-0.95$ ($\Delta_L=-20$ and $\delta=\Delta_R$ in all panels). The solid lines represent the spinon   light cones, and the dashed ones correspond to those associated with the bound state excitations.}
  \label{JZ_negativo}
\end{figure}

To strengthen the idea that the two light cones can be related to spinon and bound state excitations, we show  in Figure~\ref{JZ_negativo} color maps of the magnetization for different values of $\Delta_R<0$. 
For $\Delta_R>\Delta^*$ the maximum bound state velocity is equal to the spinon velocity, so we only see one light cone, as shown in panel a), for $\Delta_R=-0.25$. As we decrease $\Delta_R$, the bound state velocity becomes larger than the spinon one and we see oscillations outside the spinon   light cone, which form a second light cone [dashed lines in panels b), c), and d)] with a velocity that agrees well with the one obtained by Eqs. (\ref{disp}) and (\ref{velb}).



\subsection{Asymptotic behavior }\label{secB}

Let us now analyze the  asymptotic long-time state of the system in comparison with   the ground state of the  Hamiltonian after the quench, $H(t>0)$ in Eq.~(\ref{hamiltonian}), which we refer to as the equilibrium state.
%
%
%

The spin profile in the equilibrium state for the right chain with $\Delta_R=0.5$ is illustrated in the last plot of Figure~\ref{cortes_05}. Clearly,  there  are Friedel-type oscillations induced by the effective magnetic field at the boundary with  the ferromagnetic chain. The amplitude of these oscillations decay as a power law  with the distance  from the interface \cite{EggerPRL1995,FabrizioPRB1995,KitanineJSTAT2008}. The same qualitative behavior is observed  in the non-equilibrium case
deep inside the light cone, i.e., for $N_L+i\ll N_L+v_st\ll N_L+N_R$.

For a quantitative analysis, we define the distance to equilibrium as the difference of magnetization with respect to the equilibrium state
\begin{equation}
 DSz(i,t)=\frac{|Sz(i,t)-Sz_{eq}(i)|}{|Sz_{eq}(i)|},
\end{equation}
where $Sz_{eq}(i)$ is the magnetization of $i$-th site in the equilibrium state. 

\begin{figure}[t]
 \centering
 \includegraphics[scale=0.45]{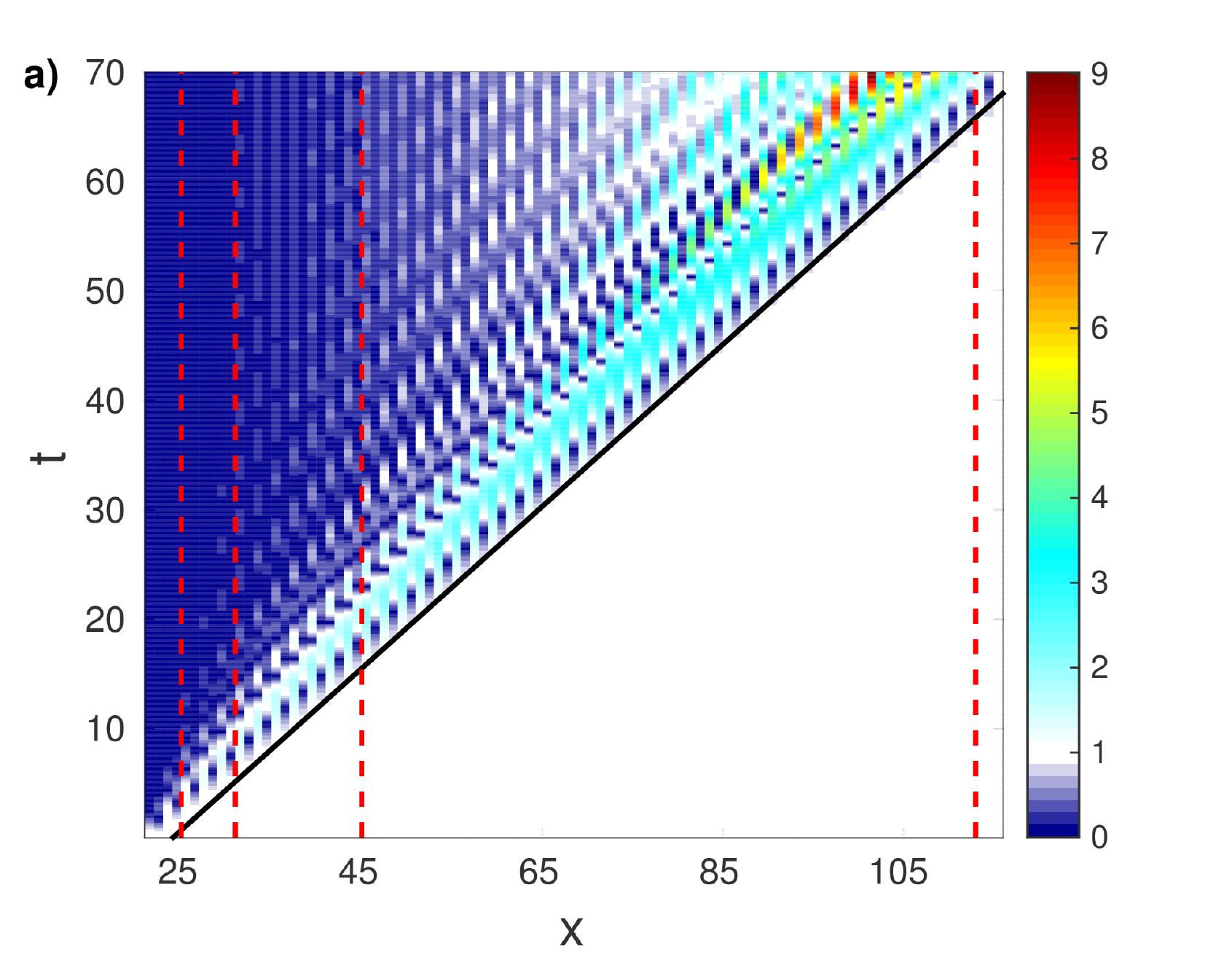}
 \includegraphics[scale=0.45]{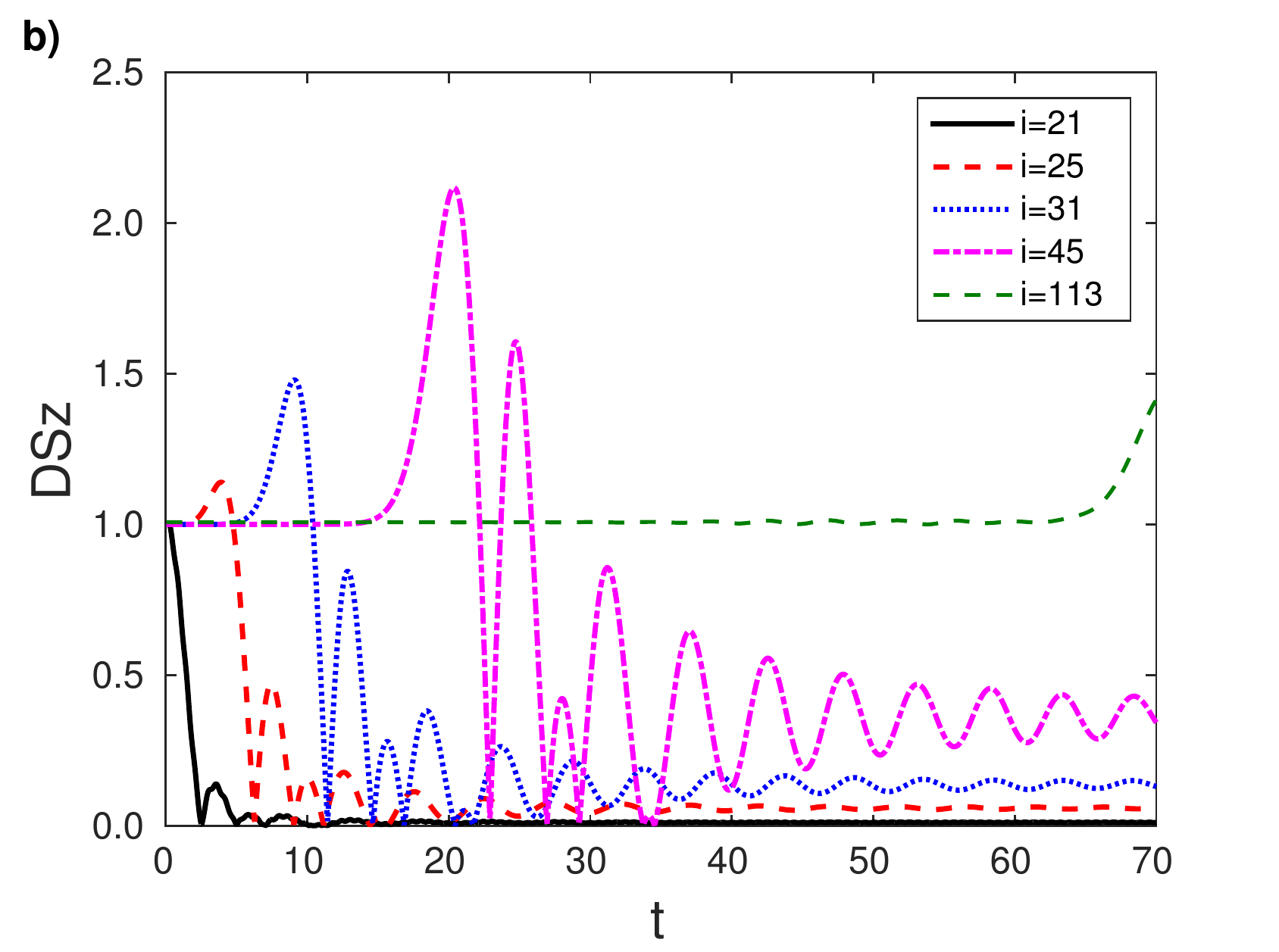}
 \caption{(Color online) a) Color map of $DSz$. The dotted lines indicate the cuts shown in panel b); the solid curve corresponds to the spinon excitation light cone boundary defined in Figure \ref{cortes_05}. b) Distance to the equilibrium state (see definition in the text) for sites $i=21$, $i=25$, $i=31$, $i=45$, and $i=113$ when the right chain has $N_R=100$ sites and $\Delta_R=0.5$. Other parameters used were $\Delta_L=-20$ and $\delta=\Delta_R$.}
 \label{distancia_05}
 \end{figure}

Figure \ref{distancia_05} a) shows the color map of $DSz(i,t)$. The solid straight line corresponds to the magnetization light cone   as defined in Figure \ref{cortes_05}. Note that initially the system was in an equilibrium state  with respect to the initial Hamiltonian $H(t<0)$, in which  $DSz(i,0)=1$ for all sites in the critical chain. After the  quench, we expect the local  magnetization to equilibrate at the values  corresponding to  the ground state of $H(t>0)$, in which case $DSz(i,t\to\infty)\to0$ for any fixed $i$. 

\begin{figure}[t]
\centering
 \includegraphics[width=\linewidth]{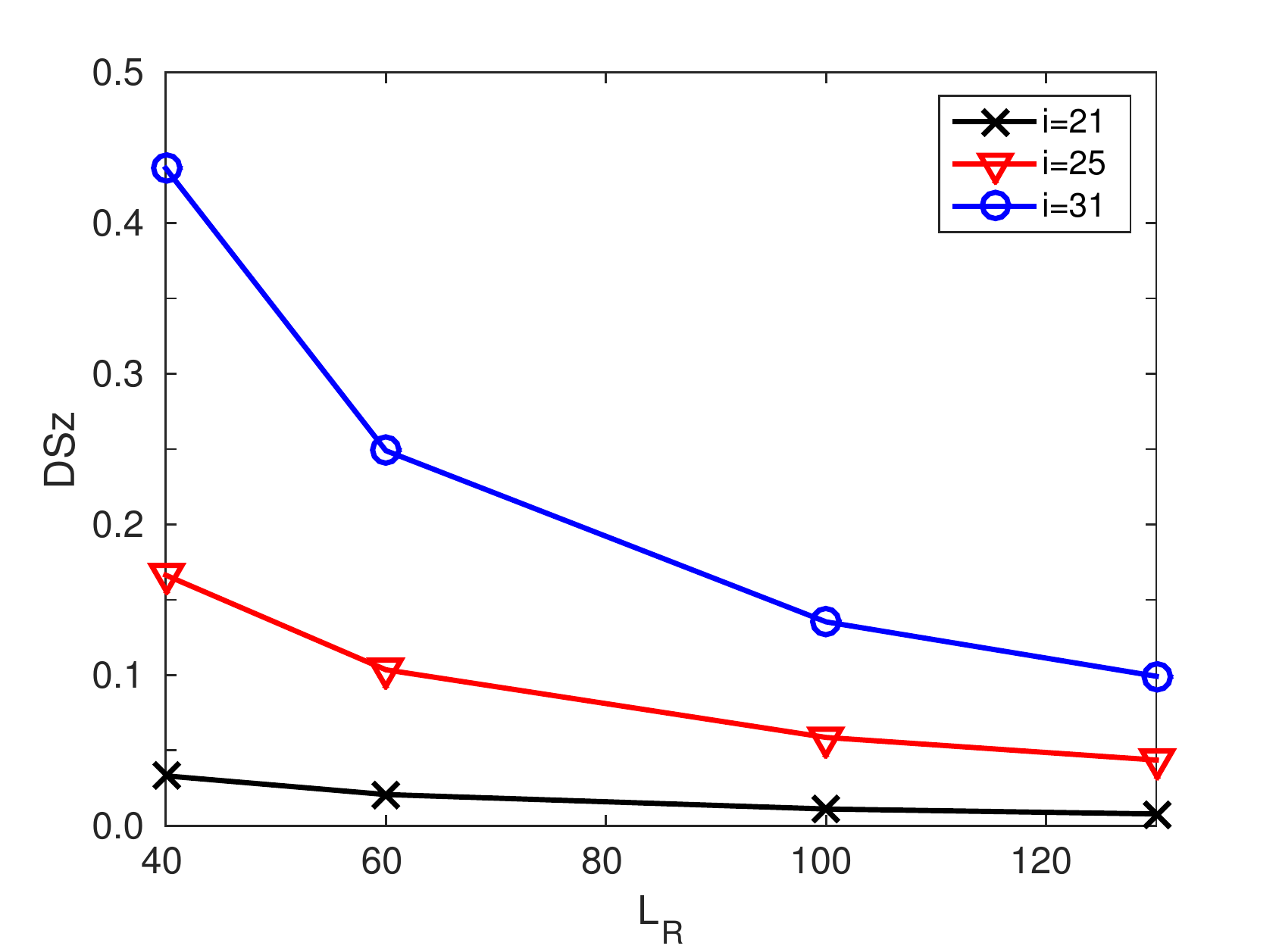}
 \caption{(Color online) Asymptotic distance to the equilibrium state for sites $i=21$, $i=25$, and $i=31$ as a function of the right chain length. Other parameters as in Figure~\ref{distancia_05}.}
 \label{tamanho}
 \end{figure}

Figure \ref{distancia_05} b) shows the distance function  $DSz(i,t)$ versus time for different sites in  a chain with $N=120$ sites ($N_L=20$ and $N_R=100$) for $\Delta_R=\delta=0.5$. For all fixed positions, the distance function shows time  oscillations  inside the light cone. However, like the equilibrium state itself,  the decay of the amplitude of the  time oscillations is  inhomogeneous. In fact,  the oscillations decay faster with time for sites near the interface than   in the bulk of the chain. A similar boundary effect has been observed in equilibrium time-dependent correlation functions of open spin chains \cite{EliensPRB2016}. Moreover, the asymptotic  value of $DSz(i,t)$ at long times  (after averaging out the oscillations) appears to approach a nonzero value that increases with the distance from the interface. We interpret this as a finite size effect. To confirm this, we analyze the distance to the equilibrium state for specific sites of the chain as function of the (right) chain
length, as shown in Figure \ref{tamanho}. This asymptotic distance was obtained through a time average over long times. 
As we increase $N_R$, the distance to equilibrium decreases,
suggesting that the whole system converges to the equilibrium state at long times only in the limit of a semi-infinite chain.

\section{Conclusion}\label{secV}

 
We have investigated the time evolution of magnetization and entanglement entropy after a local quench that connects two XXZ chains and have observed that the non-equilibrium dynamics is governed by equilibrium excitations. 
More specifically, we have connected a ferromagnetic chain to another chain in the critical phase and have seen that the quench creates excitations that propagate from the interface between them. 
When the left chain is close to the anisotropic ferromagnetic point, we see small variations in its on-site magnetization and entanglement entropy, which propagate with the maximum magnon velocity. These excitations, however, are suppressed when the chain is deep in the ferromagnetic phase, due to the strong magnon gap.   
 More interesting is the dynamics in the right chain. When it is in the critical phase with positive anisotropy parameter, the information propagates with the maximum spinon velocity,  defining a spinon light cone. When the anisotropy parameter is negative, we have observed a second light cone related to other type of excitations, the spin-wave bound states or strings, which arise in the zero magnetization subspace.

As our local quench protocol gives a small amount of energy to the system, we expect the chain to asymptotically go to the equilibrium state. In a finite system, we observe that this happens in a non-homogeneous way: the sites close to the interface are more influenced by the quench than the ones far away from it, even for long times.

We thank CNPq, CAPES, and FAPEMIG for financial support. We acknowledge Eduardo Mascarenhas and
Marcelo Fran\c{c}a Santos for useful discussions.

\newpage
 \bibliography{refer}

 \newpage

 \onecolumngrid

 \renewcommand{\baselinestretch}{1.5}


\end{document}